\RequirePackage{fix-cm}
\documentclass[smallextended]{svjour3}       % onecolumn (second format)
\smartqed  % flush right qed marks, e.g. at end of proof
\usepackage{graphicx,,bm,amsmath,placeins,color,comment}
\usepackage[a4paper,bottom=.8 in]{geometry}
\usepackage{wrapfig,lipsum,booktabs}
\usepackage{capt-of}
\usepackage{float,color}
\usepackage{amsmath,amssymb}
\usepackage{adjustbox}
\usepackage{epstopdf,placeins,cite}
\usepackage{caption}
\usepackage{subfigure}
\usepackage[colorlinks=true,linkcolor=blue]{hyperref}

\hypersetup{
	colorlinks=true,    % false: boxed links; true: colored links
	linkcolor=blue,       % color of internal links
	citecolor=blue,       % color of links to bibliography
	filecolor=red,        % color of file links
	urlcolor=blue,        % color of external links
	linktoc=page          % only page is linked
}
\begin{document}
%\begin{comment}
\title{Band Tuning of Phosphorene Semiconductor via Floquet Theory
%Quantum Floquet Oscillations in Phosphorene   %\thanks{Grants or other notes
}


\author{Km Arti Mishra \and Almas \and Upendra Kumar  %etc.
}


\institute{         %  \\
	Km Arti Mishra \at
    Department of Electrical Engineering, Faculty of Engineering and Technology, Rama University, Kanpur, UP, India-209217
	\and
	Almas \at
	Department of Electrical Engineering, Faculty of Engineering and Technology, Rama University, Kanpur, UP, India-209217
	\and
	U. Kumar \at
	Department of Physics,  Indian Institute of Technology  Guwahati,  Guwahati, Assam, India-781039\\
	\email{k.upendra@iitg.ac.in, upendraawasthi88@gmail.com}  
	}

\date{Received: date / Accepted: date}

\maketitle

\begin{abstract}
Graphene and phosphorene are monolayer of graphite and phosphorous, respectively. Graphene is  completely relativistic (Dirac) fermionic system, but phosphorene is  pseudorelativistic fermionic system. In phosphorene, electronic spectrum of phosphorene has a Dirac like (linear) band in one direction and Schr${\ddot{\mbox{o}}}$dinger like (parabolic) band in other direction. Conventional Rabi oscillations are studied by using rotating wave approximation in resonance case. The Floquet theory is an alternative way of study Rabi oscillations in off-resonance case and dominating in case of low energy physics. In this article, the nonlinear optical response of graphene and phosphorene studied under intense applied quantized electromagnetic field via Floquet theory. The Bloch-Siegert shift is observed for graphene and phosphorene. A numerical model is applied for justifying the role of anisotropy in phosphorene. Therefore, the Floquet theory can be utilized to characterize the different fermionic systems.           
\end{abstract}
\newpage
%\begin{center}
%%\section*{\huge Date-31 Oct 18- 11 Dec 18}
%\end{center}
%\end{comment}
\section{Introduction}
\noindent
The rise of many two-dimensional materials like graphene, phosphorene, silicene, etc. become a matter of curiosity for the entire field of photonics and opto-electronics \cite{li2017light}. There are many crucial and significant physical properties in graphene due to its unique electronic spectrum of massless Dirac like particles {\cite{novoselov2004electric,neto2009electronic}}. But graphene is a zero-gap semiconductor, so it has very limited applications in electronic devices without significant strain-engineering {\cite{ni2008uniaxial1}} or physical modification in the morphology {\cite{han2007energy1}}. The monolayer of black phosphorous (BP), is known as phosphorene (2D allotrope of phosphorous)\cite{liu2014phosphorene}.  There is highly anisotropic nature in phosphorene. Its band structure is quite different than other 2D materials, electronic spectrum of phosphorene has a Dirac like (linear) band in one direction and Schr${\ddot{\mbox{o}}}$dinger like (parabolic) band in other direction \cite{ezawa2015highly}. Phosphorene possess a finite band gap in its electronic spectrum \cite{guo2015from} in contrast to graphene. Phosphorene becomes a direct band gap semiconductor \cite{guo2015from} in presence of visible region of electromagnetic spectrum. So it will be highly useful in electronic devices operating in the visible region of the electromagnetic spectra like LEDs and solar cells \cite{yang2015optical}. Due to very high hole mobility, phosphorene can be utilized in P-type device materials \cite{liu2014phosphorene}. The advantage of phosphorene in compare of graphene, is the presence of tunable direct band gap, so there is tuning of nonlinear optical response of phosphorene by applying an external field \cite{li2017tunable}. There are many specific properties in phosphorene like electrical, thermal and optical anisotropy, can be utilised in device fabrication such as transparent saturable absorbers, fast photo-conductive switch and low noise photodetectors \cite{viti2018photonic}.      

Optical properties studied in the context of graphene are, optical conductivity, optical Stark effect and Rabi oscillations \cite{eberly1975optical,haug2009quantum,gerry4introductory}. Many experiments have been performed  to study optical properties, such as  optical Stark effect \cite{haug2009quantum}, optical conductivity, Rabi oscillations \cite{haug2009quantum,gerry4introductory,ni2007graphene,mandel1995optical,boyd2008nonlinear}, universal optical conductance \cite{lee1993localized,ludwig1994integer,ziegler1998delocalization}, measurement of fine structure constant \cite{nair2008fine}, four wave mixing \cite{haug2009quantum} and incoherent optical properties like optical dephasing \cite{boyd2008nonlinear}, relaxation of charge carriers, both inter-band and intra-band in graphene and graphene based systems on various substrate has been reported experimentally by pump-probe technique \cite{kumar2009femtosecond,breusing2011ultrafast,dawlaty2008measurement,shang2010femtosecond,george2008ultrafast,ruzicka2010femtosecond}.
There is presence of tunable band gap in electronic spectrum of BP, so electronic and optical properties of BP can be changed drastically \cite{wang2016optical}. There are many unique optical properties in the monolayer of BP \cite{wang2015highly} such as a large third-order nonlinear optical susceptibility of about $10^{-19} \mathrm{m^2/V^2}$, and the measured fast relaxation time is $0.13 ps$ \cite{wang2016optical,miao2017ultrafast,Margulis2017coherent}. By controlling size of BP, its nonlinear optical response can be adjusted and a new way to develop electronic and optoelectronic devices \cite{xu2017size,pedersen2017nonlinear} produced. Therefore, the optical properties (linear and nonlinear) of BP and graphene becomes a matter of curiosity for materials scientists and gives motivation to research in the field of phosphorene nonlinear optics.

If there is a cyclic change of energy between a two-level quantum system and deriving field, some oscillations produced, known as Rabi oscillations \cite{rabi1937space}. There is a lot of study has been performed on Rabi oscillations in  conventional semiconductors by using rotating wave approximation (RWA) \cite{allen1975optical,lindberg1988effective} and RWA is valid only in case of resonance. In off resonance case, there is a new type of oscillation found by application of Floquet theory \cite{oka2009photovoltaic,lindner2011floquet,kitagawa2011transport,inoue2010j,dora2012optically}. An {\it{off resonant}} light frequency applied for any electron transition in  
Floquet theory. Therefore, there is no directly excitement of electrons from light but effectively modifies the electronic band structures via virtual photon absorption processes \cite{ezawa2013photoinduced}. So Floquet theory becomes an alternative of RWA in the off-resonance case and recently applied in Dirac fermionic systems \cite{oka2009photovoltaic,lindner2011floquet,kitagawa2011transport,inoue2010j,dora2012optically}. The Floquet theory has been studied with the name of asymptotic rotating wave approximation (ARWA) by Enam {\it et al.} \cite{kumar2012crossover}. From fig. (1) of Enam {\it et al.} \cite{kumar2012crossover}, it can be explicitly seen that, the Floquet theory dominates in case of low energy physics. 

There is shift in resonance condition of RWA, know as Bloch-Siegert shift (BSS) \cite{bloch1940magnetic}, comes due to considering counter-rotating term. Such shift is very important for characterizing the amplitude, homogeneity of the proton-decoupling field and monitoring the probe performance \cite{vierkotter1996applications}. It is also found in 
a strongly driven classical two-level systems by  Beijersbergen  {\it{et al.}} \cite{beijersbergen1992multiphoton}. In graphene, BSS becomes important in case of the next nearest neighbour hopping or inclusion of Rashba spin-orbit interaction \cite{kumar2014band}. The BSS in phosphorene comes due to puckered crystal structure \cite{fukuoka2015electronic,kumar2019anisotropic}. BSS has been well studied in presence of classical \cite{shirley1965solution,bloch1940magnetic} and quantized fields \cite{stenholm1972saturation,hannaford1973analytical,cohen1973quantum}. So motivation of this work to study BSS in phosphorene and graphene, when external field is considered in quantized form.

When, there is interaction between an isolated two-level atom and a  single mode quantized electromagnetic field in a lossless cavity, Jaynes-Cummings  model \cite{jaynes1963comparison,yoo1985dynamical,Cummings} comes in picture for explanation of such phenomenon. Jaynes-Cummings  model is exactly soluble in the rotating wave approximation. Jaynes-Cummings  model has already been used for explaining collapse-revival phenomena \cite{eberly1980periodic,narozhny1981coherence,yoo1981non}. The periodic recurrence of the quantum wave function from its original form during the time evolution is known as collapse-revival, it has already been predicted by theoretical \cite{ficek2014quantum} and  experimental \cite{narozhny1981coherence} manners. The oscillations of collapse-revival, decays rapidly at short times,  but periodically regenerates to large amplitudes on a longer time scale \cite{vela2005coherent,torosov2015mixed}.        

In this article, the bands of phosphorene has been tuned via application of Floquet theory. There is presence of anisotropy in phosphorene bands, so the role of anisotropy is described in various  phenomenon like Bloch-Siegert shift, collapse-revival spectra and Floquet oscillations. There is also numerical justification of anisotropy in Floquet theory. Therefore, intrinsic anisotropy of phosphorene has major physical significance and becomes important in modern physics. The results of phosphorene compared with graphene wherever required.
\section{Collapse-Revival Spectra of Phosphorene and Graphene}
The low energy Hamiltonian of phosphorene is  $H=\left(u p_{y}^2+m\right)\sigma_{x}+v_{F} p_{x}\sigma_{y}$ \cite{ezawa2015highly}. Here $\sigma$ is the Pauli matrices, $v_{F}$ is fermi velocity, $u$ is effective mass, $m$ is the gap acting as the Dirac mass, $p_{x}$, $p_{y}$ is the $x$ and $y$ components of momentum. The low energy Hamiltonian of phosphorene in presence of vector potential ${\bf{A}}(t)$ in second quantized form is
\small
\begin{align}
H&=&c_A^{\dagger}\left[ \left(up^2_{y}+m\right)-iv_{F} p_{x}\right]c_B +
c_B^{\dagger
}\left[\left( up^2_{y}+m\right)+iv_{F} p_{x}\right]c_A+e^{ -i \omega t}\left[c_A^{\dagger} \left\{ \left(-i\frac{2up_{y}}{2v_{F}}\lambda b \right)+\frac{i}{2}\lambda b \right\} c_B\right. \nonumber\\&&
+ \left.
c_B^{\dagger}\left\{\left(-i\frac{2up_{y}}{2v_{F}}\lambda b \mbox{   }\right)-\frac{i}{2}\lambda b\right\}c_A
\right]
+e^{ i \omega t}\left[ c_A^{\dagger}\left\{i\left(\frac{2up_{y}}{2v_{F}}\lambda b^{\dagger} \right)+\frac{i}{2}\lambda b^{\dagger}\right\}c_B \right.  \nonumber\\&& + \left.
c_B^{\dagger}\left\{\left(i\frac{2up_{y}}{2v_{F}}\lambda b^{\dagger} \mbox{   }\right)-\frac{i}{2}\lambda b^{\dagger}\right\}c_A
\right]
\label{JCHP}
\end{align}
\normalsize
Here $ A, B$ represents either spin up or spin down  and $c\mbox{  }(c^{\dagger})$ for the annihilation (creation) operator. The vector is considered as ${\bf{A}}(t)=\mathrm{Re}({\bf{A_{0}}} e^{-i\omega t})$ i.e. for circular polarization. The Floquet oscillations are present in only case of circular polarization \cite{kumar2014quantum}. The coupling constant $\lambda$ is defined as $-\frac{ev}{c}{\bf{A_{0}}}=\lambda b$, $ [b,b^{\dagger}] = 1 $ are the photon operators. The Hamiltonian model (eq.(\ref{JCHP})) becomes analogous to Jaynes-Cummings model by using identifications $\sigma_{+}=\sigma_{x}+i\sigma_{y} = c^{\dagger}_{B} c_{A}$, $\sigma_{-}=\sigma_{x}-i\sigma_{y} = c^{\dagger}_{A}c_{B}$ and $\sigma_{z} =(c_A^{\dagger} c_A-c_B^{\dagger}c_B)$. There is unitary transformation on the photons i.e. replace $ b $ with $ b e^{ i \omega t }$. For making above model (eq.(\ref{JCHP})) simpler, there is consideration of  one electron hoping. Therefore, the transition states has form $\langle0,1,n+1|\phi(t)\rangle$ and $\langle1,0,n+1|\phi(t)\rangle$, with non-zero amplitudes.  $n$ is number of the photon, which is very large, so it can be considered $ n \approx   n+1 $. Therefore, 
the energy-eigenvalue equation $i\hbar\frac{\partial}{\partial t}|\phi(t)\rangle=H |\phi(t)\rangle$ of phosphorene in matrix form (setting $\hbar=1$)
\begin{align}&&i\hbar\frac{\partial}{\partial_t}\begin{bmatrix}
\langle 0,1,n+1|\phi \rangle \\ \langle 1,0,n+1|\phi \rangle
\end{bmatrix}=\begin{bmatrix}
0& \left[\left\{ up^2_{y}+m\right\}+iv_{F} p_{x}\right]\\ \left[ \left\{ up^2_{y}+m\right\}-iv_{F} p_{x}\right] &0
\end{bmatrix}\begin{bmatrix}
\langle 0,1,n+1|\phi \rangle \\ \langle 1,0,n+1|\phi \rangle
\end{bmatrix} \nonumber \\&&
+e^{ -i \omega t}\begin{bmatrix}0&
\left[ \left\{-i\frac{2up_{y}}{2v_{F}}\lambda  \mbox{   }\right\}-\frac{i}{2}\lambda \right]\;\sqrt{n+1}\\
\left[ \left\{-i\frac{2up_{y}}{2v_{F}}\lambda \right\}+\frac{i}{2}\lambda  \right]\;\sqrt{n+1}&0
\end{bmatrix}\begin{bmatrix}
\langle 0,1,n+1|\phi \rangle \\ \langle 1,0,n+1|\phi \rangle
\end{bmatrix} \nonumber \\&& +e^{ i \omega t}\begin{bmatrix}0&
\left[ \left\{i\frac{2up_{y}}{2v_{F}}\lambda  \mbox{   }\right\}-\frac{i}{2}\lambda  \right]\;\sqrt{n+1}\\
\left[ \left\{i\frac{2up_{y}}{2v_{F}}\lambda  \right\}+\frac{i}{2}\lambda   \right]\;\sqrt{n+1}&0
\end{bmatrix}\begin{bmatrix}
\langle 0,1,n+1|\phi \rangle \\ \langle 1,0,n+1|\phi \rangle
\end{bmatrix}
\label{matrixP}
\end{align}
Similarly, the low energy Hamiltonian of graphene $H=v_{F}(\sigma_{x} p_{x}+\sigma_{y} p_{y})$ \cite{neto2009electronic} in the presence of vector potential ${\bf{A}}(t)$ in second quantized form  
\begin{align}
H &=&  c^{\dagger}_{A}\mbox{  } v_{F}( p_{x}-i p_{y})\mbox{  } c_{B} + c^{\dagger}_{B}\mbox{  } v_{F}( p_{x}+i p_{y})\mbox{  } c_{A}
+  \lambda \mbox{   } c_B^{\dagger}c_A\mbox{  } b \mbox{  }e^{ i \omega t } + \lambda\mbox{  }c_A^{\dagger}c_B
\mbox{  }b^{\dagger}\; e^{ -i \omega t }.
\label{JCHG}
\end{align}
Therefore, matrix form of the energy-eigenvalue equation $i\hbar\frac{\partial}{\partial t}|\phi(t)\rangle=H |\phi(t)\rangle$ of graphene (setting $\hbar=1$)
\begin{eqnarray}
	i\frac{\partial}{\partial_{t}}\left[
	\begin{array}{cc}
	\langle 0,1,n+1\mbox{  }|\phi(t) \rangle\\
	\langle 1,0,n+1\mbox{  }|\phi(t) \rangle\\
	\end{array}\right]
	&=&\left[\begin{array}{ccc}
	0& v_{F}(p_{x}+ip_{y})\\
	v_{F}(p_{x}-ip_{y})&0\\
	\end{array}\right]\left[\begin{array}{cc}
	\langle 0,1,n+1\mbox{  }|\phi(t) \rangle\\
	\langle 1,0,n+1\mbox{  }|\phi(t) \rangle\\
	\end{array}\right] \nonumber \\&&+e^{- i \omega t } \left[\begin{array}{cc}
	0&0\\
	\lambda \sqrt{n+1}&0\\
	\end{array}\right]
	\left[\begin{array}{cc}
	\langle 0,1,n+1\mbox{  }|\phi(t) \rangle\\
	\langle 1,0,n+1\mbox{  }|\phi(t) \rangle\\
	\end{array}\right]\nonumber\\&&+e^{ i \omega t }\left[\begin{array}{cc}
	0&\lambda \sqrt{n+1}\\
	0&0\\
	\end{array}\right]
	\left[\begin{array}{cc}
	\langle 0,1,n+1\mbox{  }|\phi(t) \rangle\\
	\langle 1,0,n+1\mbox{  }|\phi(t) \rangle\\
	\end{array}\right].
	\label{matrixG}
\end{eqnarray}

\subsection{Rotating wave approximation (RWA)}\label{RWA-appox}
\subsubsection{Phosphorene}\label{RWAP}
With help well known RWA \cite{haug2009quantum}, the eq.(\ref{matrixP}) can be solved. In such approximation, the rapidly oscillating terms of the effective Hamiltonian are neglected. First, the first matrix of eq.(\ref{matrixP}) is diagonalize by using unitary transformation      
\begin{align}\begin{bmatrix}
\langle 0,1,n+1|\phi \rangle \\ \langle 1,0,n+1|\phi \rangle
\end{bmatrix}
=\left(
\begin{array}{cc}
-\frac{u p_{y}^2+m+i p_{x} v_{F}}{\beta} & \frac{u p_{y}^2+m+i p_{x} v_{F}}{\beta} \\
1 & 1 \\
\end{array}
\right)\begin{bmatrix}
\langle0,1,n+1|\phi \rangle_{1} \\ \langle 1,0,n+1|\phi_{1} \rangle_{1}
\end{bmatrix}
\end{align}
Here, ${ \beta =\sqrt{\left(u p_{y}^2+m\right)^2+p_{x}^2 v_{F}^2}}$. Using above transformation, eq.(\ref{matrixP}) becomes
\small
\begin{align}\hspace{-1.5cm} &&i\hbar\frac{\partial}{\partial_t}\begin{bmatrix}
\langle0,1,n+1|\phi \rangle_{1} \\ \langle 1,0,n+1|\phi_{1} \rangle_{1}
\end{bmatrix}= 
\left(
\begin{array}{cc}
-\beta & 0 \\
0 & \beta \\
\end{array}
\right)
\begin{bmatrix}
\langle0,1,n+1|\phi \rangle_{1} \\ \langle 1,0,n+1|\phi_{1} \rangle_{1}
\end{bmatrix} +e^{ -i \omega t}\sqrt{n+1}\;\lambda\left(
\begin{array}{cc}
\frac{ \left[p_{x} v_{F}^2+2 i p_{y} u \left(u p_{y}^2+m\right)\right]  }{2 v_{F} \beta} & \frac{i  \left[m+p_{y} (p_{y}-2 i p_{x}) u\right]  }{2 \beta} \\
-\frac{i  \left[m+p_{y} (p_{y}-2 i p_{x}) u\right] }{2 \beta} & -\frac{i  \left[2 p_{y} u \left(u p_{y}^2+m\right)-i p_{x} v_{F}^2\right]  }{2 v_{F} \beta} \\
\end{array}
\right)\nonumber \\&&
\begin{bmatrix}
\langle0,1,n+1|\phi \rangle_{1} \\ \langle 1,0,n+1|\phi_{1} \rangle_{1}
\end{bmatrix}  +e^{ i \omega t}\sqrt{n+1}\;\lambda\left(
\begin{array}{cc}
\frac{ \left[p_{x} v_{F}^2-2 i p_{y} u \left(u p_{y}^2+m\right)\right] }{2 v_{F} \beta} & \frac{i \left[m+p_{y} (2 i p_{x}+p_{y}) u\right]) }{2 \beta} \\
-\frac{i \left[m+p_{y} (2 i p_{x}+p_{y}) u\right] }{2 \beta} & \frac{i  \left[i p_{x} v_{F}^2+2 p_{y} u \left(u p_{y}^2+m\right)\right] }{2 v_{F} \beta} \\
\end{array}
\right)
\begin{bmatrix}
\langle0,1,n+1|\phi \rangle_{1} \\ \langle 1,0,n+1|\phi_{1} \rangle_{1}
\end{bmatrix}
\end{align}
\normalsize
Now again using new transformation  $\langle0,1,n+1|\phi \rangle_{1}=e^{i t \beta }\langle0,1,n+1|\phi \rangle_{2}$ and $\langle1,0,n+1|\phi \rangle_{1}=e^{-i t \beta }\langle1,0,n+1|\phi \rangle_{2}$ and leaving the counter-rotating term (rapidly varying term), the final equation of RWA has form
\begin{subequations}
\begin{align}
i\frac{\partial}{\partial_t} \langle0,1,n+1|\phi \rangle_{2}=\frac{i \left[m+p_{y} (2 i p_{x}+p_{y}) u\right] \sqrt{n+1}\;\lambda}{2 \beta}\;e^{ i (\omega-2\beta) t}\langle 1,0,n+1|\phi_{1} \rangle_{2}
\label{RWA1}
\end{align}
\begin{align}
i \frac{\partial}{\partial_t}\langle 1,0,n+1|\phi_{1} \rangle_{2}= \;\frac{i  \left[m+p_{y} (p_{y}-2 i p_{x}) u\right] \sqrt{n+1}\;\lambda}{2 \beta}e^{ -i (\omega-2\beta) t}\langle0,1,n+1|\phi \rangle_{2}
\label{RWA2}
\end{align}
\end{subequations}
Applying initial condition $\langle0,1,n+1|\phi \rangle_{2}=0$ and $\langle 1,0,n+1|\phi_{1} \rangle_{2}=1$ and solving above eq.(\ref{RWA1}) and eq.(\ref{RWA2}), the probability amplitude of wave-function $\langle 1,0,n+1|\phi_{1} \rangle_{2}$ is   
\begin{align}P_{2P-RWA}(t)=|\langle 1,0,n+1|\phi_{1} \rangle_{2}|^2=\cos^2\left(\frac{t}{2} \Omega_{\mathrm{RWA-P}}(n)\right).
\end{align}
Where $\Omega_{\mathrm{RWA-P}}(n)=\sqrt{\Delta_{P}^2+(n+1) \lambda^2\gamma}$ is conventional Rabi frequency of phosphorene, $\Delta_{P}=(\omega-2\beta)$ is detuning parameter and $\gamma=\left[(m+p^2 u \sin ^2\theta )^2+p^4 u^2 \sin ^22 \theta \right]/\beta^2$ by considering momentum vector in polar form i.e. $p_x=p \cos \theta $, $p_y=p \sin \theta$ and $\hbar=1$. In RWA, detuning $\Delta_{P}$ becomes zero. 
\begin{figure}[h]
	\centering
	\subfigure[]{\includegraphics[width=70mm,height=65mm]{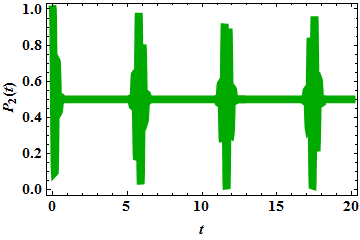}}
	\subfigure[]{\includegraphics[width=70mm,height=65mm]{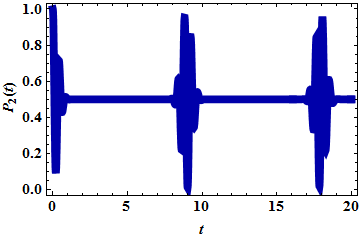}}\\
	\caption{\small (Color online) The Collapse and revival phenomenon of Rabi oscillations in phosphorene for different value of wavevector angle {\bf(a)} $\theta=\frac{\pi}{4}$ {\bf(b)} $\theta=0$. The plot between the probability of state $\langle 0,1,n+1\mbox{  }|\phi(t) \rangle$ and time. For plotting, we considered the mean number of photon $\langle n \rangle =20$, $\lambda=1$, $u=1$, $m=1$, $v_{F}=1$ $p=1$ and  $\omega=1$. Time is in the unit of $\lambda^{-1}$.}
	\label{collapse_revival_RWA}
\end{figure}
%\FloatBarrier
\noindent Taking initial conditions in quantized form i.e. values of the probability amplitudes are
$\langle 0,1,n\mbox{  }|\phi(t) \rangle=\langle n|\alpha\rangle\; \mbox{and} \;\langle 1,0,n\mbox{  }|\phi(t) \rangle=0
$, which gives
\begin{align}|\langle 0,1,n\mbox{  }|\phi(t) \rangle|^2=\frac{|\alpha|^{2n}}{n!}e^{-|\alpha|^2}.
\end{align}
Where $\alpha$ is a complex number and $|\alpha|^{2}=\langle n \rangle$ is the mean number of photons in the cavity field. The population $P_{2}(t)$ comes in form 
\begin{align}
P_{2P-RWA}(t)=\sum_{n}\frac{\langle n \rangle^{n}}{n!}e^{-\langle n \rangle}\cos^2\left(\frac{t}{2} \Omega_{\mathrm{RWA-P}}(n)\right).
\label{P2-RWA}
\end{align}
The conventional Rabi frequency start to spread after coming Poisson distribution of the photon number $n$ in the picture. Due to Poisson distribution, there is dephasing in Rabi oscillations and collapse after some time $t$. There is revival of the collapsed comes due to the phases of oscillation of neighbouring terms in eq.(\ref{P2-RWA}) differ by the factor $2\pi$ \cite{dung1990collapses}. The collapse and revival in Rabi oscillation can be explicitly seen by plotting $P_{2P-RWA}(t)$ with respect to time $t$ [eq.(\ref{P2-RWA})]. Therefore, conventional Rabi frequency contains anisotropic nature, depicted in fig.(\ref{collapse_revival_RWA}). $P_{2P-RWA}(t$ has different value of amplitudes for different value of wave-vector angle $\theta$ [fig.(\ref{collapse_revival_RWA})]. The detailed analysis about collapse and revival phenomenon can be seen in the book of Scully et al. \cite{scully1999quantum}, the expression of  collapse and revival time is derived as \cite{kumar2014quantum}     
\begin{align}t_{\mathrm{col}}(\bar{n})=\sqrt{\frac{2Log(10)}{\bar{n}}}\frac{\bar{n}}{\Omega_{RWA}(\bar{n})},\quad t_{\mathrm{rev}}(\bar{n})=\frac{2\pi\bar{n} }{\Omega_{RWA}(\bar{n})}.
\end{align}
\begin{figure}[H]
	\centering
	\subfigure[]{\includegraphics[width=70mm,height=65mm]{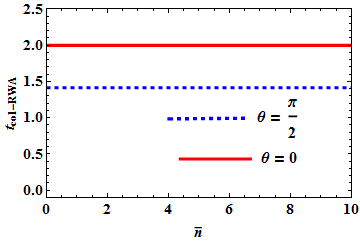}}
	\subfigure[]{\includegraphics[width=70mm,height=65mm]{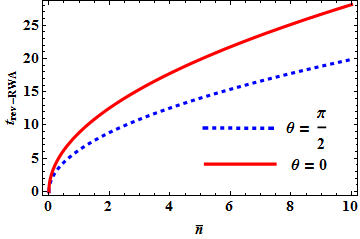}}\\
	\caption{\small (Color online) For Rabi oscillations of phosphorene {\bf(a)} collapse time, when wave vector angle $\theta=\frac{\pi}{2}$ and $0$. {\bf(b)}  revival time, when wave vector angle $\theta=\frac{\pi}{2}$ and $0$. For plotting, we considered $\lambda=1$ and $\omega=1$. Time is considered in the unit of $\lambda^{-1}$.}
	\label{collapse_revival_time_RWA}
\end{figure}
\FloatBarrier
\noindent 
Therefore, in case of phosphorene final expression of of collapse and revival time for conventional Rabi frequency is  
\begin{align}t_{\mathrm{col-RWA}}(\bar{n})=\sqrt{2Log(10)}\frac{1}{\lambda \sqrt{\gamma}},\quad t_{\mathrm{rev-RWA}}(\bar{n})=\frac{2\pi \sqrt{\bar{n}}}{\lambda \sqrt{\gamma}}.
\label{colrev-RWA}
\end{align}
\noindent 
The plot of collapse and revival time of phosphorene conventional Rabi frequency [eq.(\ref{colrev-RWA})] is depicted in fig.(\ref{collapse_revival_time_RWA}). The $t_{\mathrm{col}}$ is not changing with the increasing number of photon[fig.\ref{collapse_revival_time_RWA}({\bfseries{a}})], on the other hand $t_{\mathrm{rev}}$ is changing in continuous way as number of photons increases [fig.\ref{collapse_revival_time_RWA}({\bfseries{b}})].  But $t_{\mathrm{col}}$ and $t_{\mathrm{rev}}$ possess an anisotropic nature in phosphorene.  
\subsubsection{Graphene}\label{RWAG}
Doing similar analogy like earlier section (\ref{RWAP}), the final equation of RWA for graphene  
\begin{subequations}
	\begin{align}
	i\frac{\partial}{\partial_t} \langle0,1,n+1|\phi \rangle_{2}=-\frac{\sqrt{n+1}\mbox{ }\lambda\;e^{-i \Delta_{G} t}v_{F}(p_{x}-i\;p_{y} )}{2v_{F} \sqrt{p_{x}^2+p_{y}^2}}\langle 1,0,n+1|\phi_{1} \rangle_{2}
	\label{RWA3}
	\end{align}
	\begin{align}
	i \frac{\partial}{\partial_t}\langle 1,0,n+1|\phi_{1} \rangle_{2}= - \frac{\sqrt{n+1}\;\lambda e^{i \Delta_{G} t}\mbox{ } v_{F}(p_{x} +i p_{y} )}{2v_{F} \sqrt{p_{x}^2+p_{y}^2}}\langle0,1,n+1|\phi \rangle_{2}
	\label{RWA4}
	\end{align}
\end{subequations}
Here $\Delta_{G}=\omega-2v_{F}\sqrt{p_{x}^2+p_{y}^2}$, which is detuning parameter in case of graphene. Applying initial condition in quantized form [like earlier section (\ref{RWAP})], the probability amplitude of wave-function $\langle 1,0,n+1|\phi_{1} \rangle_{2}$ is   
\begin{align}
P_{2G-RWA}(t)=\sum_{n}\frac{\langle n \rangle^{n}}{n!}e^{-\langle n \rangle}\cos^2\left(\frac{t}{2} \Omega_{\mathrm{RWA-G}}(n)\right).
\label{P2-RWAG}
\end{align}
Where $\Omega_{\mathrm{RWA-G}}(n)=\sqrt{\Delta_{G}^2+(n+1) \lambda^2}$ is conventional Rabi frequency of graphene. $\Delta_{G}$ is detuning parameter, becomes zero in case of RWA. The expression of  collapse and revival time \cite{kumar2014quantum} for graphene is     
\begin{align}t_{\mathrm{colRWA}}(\bar{n})=\sqrt{\frac{2Log(10)}{\bar{n}}}\frac{\bar{n}}{\lambda\sqrt{\bar{n}+1}},\quad t_{\mathrm{revRWA}}(\bar{n})=\frac{2\pi\bar{n} }{\lambda\sqrt{\bar{n}+1}}.
\label{tcolrevG}
\end{align}
\begin{figure}[h]
	\centering\subfigure[]{\includegraphics[width=70mm,height=65mm]{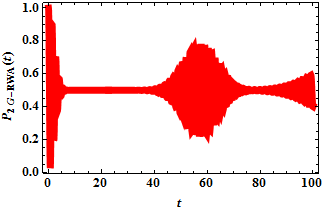}}
	\subfigure[]{\includegraphics[width=70mm,height=65mm]{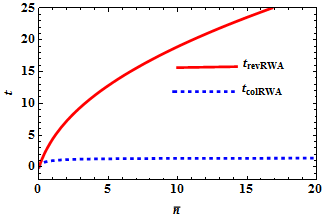}}\\
	\caption{\small (Color online) The plot {\bf(a)} showing isotropic nature of collapse-revival phenomenon in graphene and {\bf(b)} showing collapse and revival time nature for graphene. Both plot are taken in case of conventional Rabi frequency. For plotting, we considered the mean number of photon $\langle n \rangle =20$, $\lambda=1$ and $v_{F}=1$. Time is in the unit of $\lambda^{-1}$.}
	\label{collapse_revival_RWAG}
\end{figure}
\noindent The plot related with collapse-revival phenomenon [eq.(\ref{P2-RWAG})], collapse time $t_{\mathrm{col}}$ and revival time $t_{\mathrm{rev}}$ [eq.(\ref{tcolrevG})] is depicted in fig.(\ref{collapse_revival_RWAG}) for  graphene. It can be seen in fig.\ref{collapse_revival_RWAG}{\bf(b) }that $t_{\mathrm{revRWA}}$ is not varying with number of photon, on the other hand $t_{\mathrm{colRWA}}$ is changing drastically in case of graphene.
\subsection{Floquet theory approximation}\label{Floq}
\subsubsection{Phosphorene}\label{FloqP}
If the external driving frequency $(\omega)$ is nearly equal to the  particle-hole pairs $(2v|p|)$ frequencies or resonant frequencies $(\omega_R)$ of the system i.e. $\omega \approx \omega_R$ and $\omega \approx 2v|p| $,  energy eigenvalue equation ($H\psi=E\psi$) are solved by using rotating wave approximation (RWA) \cite{haug2009quantum}, described in earlier section(\ref{RWA-appox}). On the other hand, when  $\omega$ is too large compare to the Rabi frequency $\omega_R$ and  the resonant frequency of the creation particle-hole pairs i.e. $2v|p |$, $\omega \gg \omega_R$ and $\omega \gg 2v|p|$ (off-resonant case),  the Floquet approximation is applied to solve energy eigenvalue equation. In Floquet theory, the Hamiltonian is decomposed in series harmonics i.e. $H = H_0 + e^{ -i \omega t } \mbox{   }V_{+} + e^{ i \omega t } \mbox{    }V_{-}$. Similarly, writing wave-function in series harmonics form $\psi = \psi_0 + e^{ -i \omega t } \mbox{   }\psi_{+} + e^{ i \omega t } \mbox{    }\psi_{-}$. $H_0$ and $\psi_0$ related to slow parts,  on the other hand $V_{+},V_{-}$  and $\psi_{+},\psi_{-}$ related to the (coefficients of) fast parts of the full Hamiltonian and wave-function, respectively. By application of Floquet theory conditions i.e. external driving frequency $\omega$ contains larger value than band gap. Putting all these expression into energy eigenvalue equation $H\psi=E\psi$ and leaving higher harmonics i.e. order of $\frac{1}{\omega^{2}}$. Eventually, writing Hamiltonian $H$ in term of slow part only 
\begin{eqnarray}
H_{eff}=\left(H_0\mbox{   } +\frac{1}{\omega} \left[
V_{-} , \mbox{   }V_{+}\right]\right).
\label{firstorder}
\end{eqnarray}
The {\it Floquet oscillations frequency} is eigenvalues of $H_{eff}$. Therefore, by comparison of eq.(\ref{matrixP}) with $H|\phi(t)\rangle =\left( H_0 + e^{ -i \omega t } \mbox{   }V_{+} + e^{ i \omega t } \mbox{    }V_{-}\right)|\phi(t)\rangle$, the value of $H_{0}$, $V_{+}$  and $V_{-}$ will found. Therefore, Floquet energy eigenvalue equations i.e. $i\frac{\partial}{\partial_{t}}|\phi(t)\rangle=H_{eff}|\phi(t)\rangle$ have form
\begin{subequations}
\begin{align}i\hbar\frac{\partial}{\partial_t}\langle 0,1,n+1|\phi \rangle&=&\left(\frac{\frac{(n+1) (2 p_{y} u-v_{F})^2 \lambda  \lambda}{4 v^{2}_{F}}-\frac{(n+1) (2 p_{y} u+v_{F})^2 \lambda  \lambda}{4 v^{2}_{F}}}{\omega }\right)\langle 0,1,n+1|\phi \rangle \nonumber\\&&+\left(u p_{y}^2+m+i p_{x} v_{F}\right)\langle 1,0,n+1|\phi \rangle
\end{align}
\begin{align}
i\hbar\frac{\partial}{\partial_t}\langle 1,0,n+1|\phi \rangle&=&\left(\frac{\frac{(n+1) (2 p_{y} u+v_{F})^2 \lambda  \lambda}{4 v^{2}_{F}}-\frac{(n+1) (2 p_{y} u-v_{F})^2 \lambda  \lambda}{4 v^{2}_{F}}}{\omega }\right)\langle 1,0,n+1|\phi \rangle \nonumber\\&&+\left(u p_{y}^2+m-i p_{x} v_{F}\right)\langle 0,1,n+1|\phi \rangle
\end{align}
\end{subequations}
Therefore, the value of probability of state $\langle 1,0,n+1\mbox{  }|\phi(t) \rangle$ in  polar coordinate is
\begin{align}
P_{2P-Floq}(t)=|\langle 1,0,n+1\mbox{  }|\phi(t) \rangle|^2=\left[\frac{v_{F}^2 \omega ^2 \beta^2 }{v_{F}^2 \omega ^2 \beta^2+4 \lambda ^4 (n+1)^2 p^2 u^2 \sin ^2\theta }\right]
\sin ^2\left[t\Omega_{\mathrm{Floq}}(n)\right].
\end{align}
Where, $\beta =\sqrt{\left(m+p^2 u \sin ^2\theta \right)^2+p^2 v_{F}^2 \cos ^2\theta}$ and Floquet frequency 
\begin{align}
\Omega_{\mathrm{Floq-P}}(n) = \sqrt{\beta^2+\frac{4 \lambda ^4 (n+1)^2 p^2 u^2 \sin ^2\theta}{v_{F}^2 \omega ^2 }}.
\label{ARWAWP}
\end{align}
Choosing initial condition in quantized form like earlier section (\ref{RWA-appox}),
the population $P_{2}(t)$ has form
\begin{align}
P_{2P-Floq}(t)=\sum_{n}\frac{\langle n \rangle^{n}}{n!}e^{-\langle n \rangle}\left[\frac{v_{F}^2 \omega ^2 \beta^2 }{v_{F}^2 \omega ^2 \beta^2+4 \lambda ^4 (n+1)^2 p^2 u^2 \sin ^2\theta }\right]
\sin ^2\left[t\Omega_{\mathrm{Floq}}(n)\right].
\label{P2-Floq}
\end{align}
\begin{figure}[H]
\centering
\subfigure[]{\includegraphics[width=70mm,height=65mm]{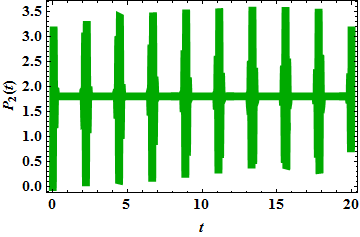}}
\subfigure[]{\includegraphics[width=70mm,height=65mm]{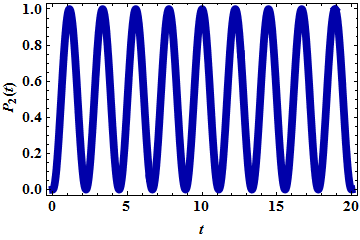}}\\
\caption{\small  (Color online) The Collapse and revival phenomenon of Floquet oscillations in phosphorene for different value of wave vector angle {\bf(a)} $\theta=\frac{\pi}{4}$ {\bf(b)} $\theta=0$. The plot between the probability of state $\langle 0,1,n+1\mbox{  }|\phi(t) \rangle$ and time. For plotting, we considered the mean number of photon $\langle n \rangle =20$, $\lambda=1$, $u=1$, $m=1$, $v_{F}=1$ $p=1$ and  $\omega=1$. Time is in the unit of $\lambda^{-1}$.}
	\label{collapse_revival_Floq}
\end{figure}
\noindent Therefore, fig.\ref{collapse_revival_Floq}{\bf(a)} showing collapse revival phenomenon of Floquet oscillations in phosphorene, when wave-vector angle $\theta=\frac{\pi}{4}$ and collapse revival phenomenon vanished in fig.\ref{collapse_revival_Floq}{\bf(b)}, when wave-vector angle $\theta=0$. Doing similar analogy like earlier section (\ref{RWA-appox}), the expression for $t_{\mathrm{col-Floq}}$ and $t_{\mathrm{rev-Floq}}$ has form
\begin{align}t_{\mathrm{col}}(\bar{n})=\sqrt{\frac{2Log(10)}{\bar{n}}}\frac{\bar{n}}{\Omega_{\mathrm{Floq}}(\bar{n})},\quad t_{\mathrm{rev}}(\bar{n})=\frac{2\pi\bar{n} }{\Omega_{\mathrm{Floq}}(\bar{n})}.
\end{align}
\noindent The plot associated with Floquet frequency collapse and revival time [eq.(\ref{ARWAWP})], given in fig.(\ref{collapse_revival_time_Flq}). From fig.(\ref{collapse_revival_time_Flq}), it can be seen that $t_{\mathrm{col-Floq}}$ and $t_{\mathrm{rev-Floq}}$  has a crucial dependency on anisotropy. For the different value of wave-vector  angle $\theta$, collapse and revival time of Floquet oscillations have drastic  changes.
\begin{figure}[H]
	\centering
	\subfigure[]{\includegraphics[width=70mm,height=65mm]{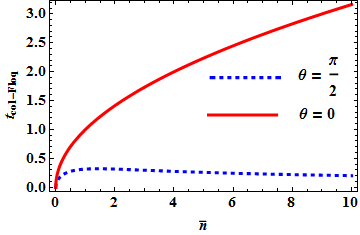}}
	\subfigure[]{\includegraphics[width=70mm,height=65mm]{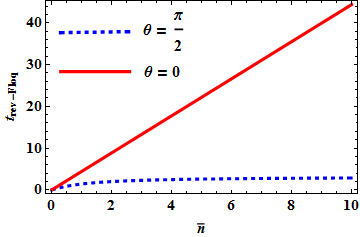}}\\
	\caption{\small (Color online)For Floquet oscillations {\bf(a)} collapse time, when wave vector angle $\theta=\frac{\pi}{2}$ and $0$. {\bf(b)}  revival time, when wave vector angle $\theta=\frac{\pi}{2}$ and $0$. For plotting, we considered $\lambda=1$ and $\omega=1$. Time is considered in the unit of $\lambda^{-1}$.}
	\label{collapse_revival_time_Flq}
\end{figure}
%%%%%%%%%%%%%%%%%%%%%%%%%%%%%%%%%%%%%
\subsubsection{Graphene}
\begin{figure}[h]
	\centering
		\subfigure[]{\includegraphics[width=70mm,height=65mm]{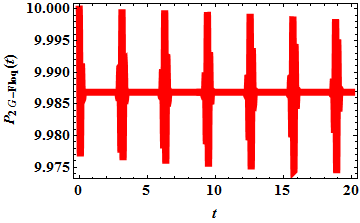}}
	\subfigure[]{\includegraphics[width=70mm,height=65mm]{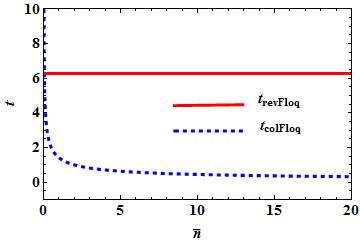}}\\
	\caption{\small (Color online) The plot {\bf(a)} showing isotropic nature of collapse-revival phenomenon in graphene and {\bf(b)} showing collapse and revival time nature for graphene. Both plot are taken in case of Floquet frequency. For plotting, we considered the mean number of photon $\langle n \rangle =20$, $\lambda=1$, $\omega=1$ and $v_{F}=1$. Time is in the unit of $\lambda^{-1}$.}
	\label{collapse_revival_FloqG}
\end{figure}

\begin{figure}[H]
	\centering
	\subfigure[]{\includegraphics[width=70mm,height=65mm]{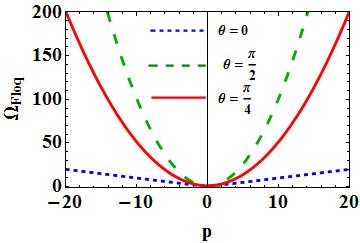}}
\subfigure[]{\includegraphics[width=70mm,height=65mm]{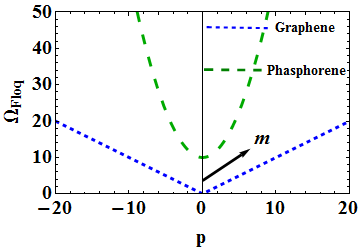}}\\
	\caption{(Colour Online) \textbf{(a)} Shows the Floquet frequency with respect to momentum in phosphorene for different values of wave vector angle $\theta$, which shows the anisotropic behaviour.    \textbf{(b)} Comparison of Floquet frequency in phosphorene $(\theta=\frac{\pi}{4})$ and graphene. For plotting, we have taken $\lambda=0.1$, $\omega=1$, $n=10$ and $v_{F}=1$. All parameters are taken in the unit of $\omega^{-1}$.}
	\label{Rel}
\end{figure}

\begin{figure}[H]
	\centering
	\subfigure[]{\includegraphics[width=70mm,height=65mm]{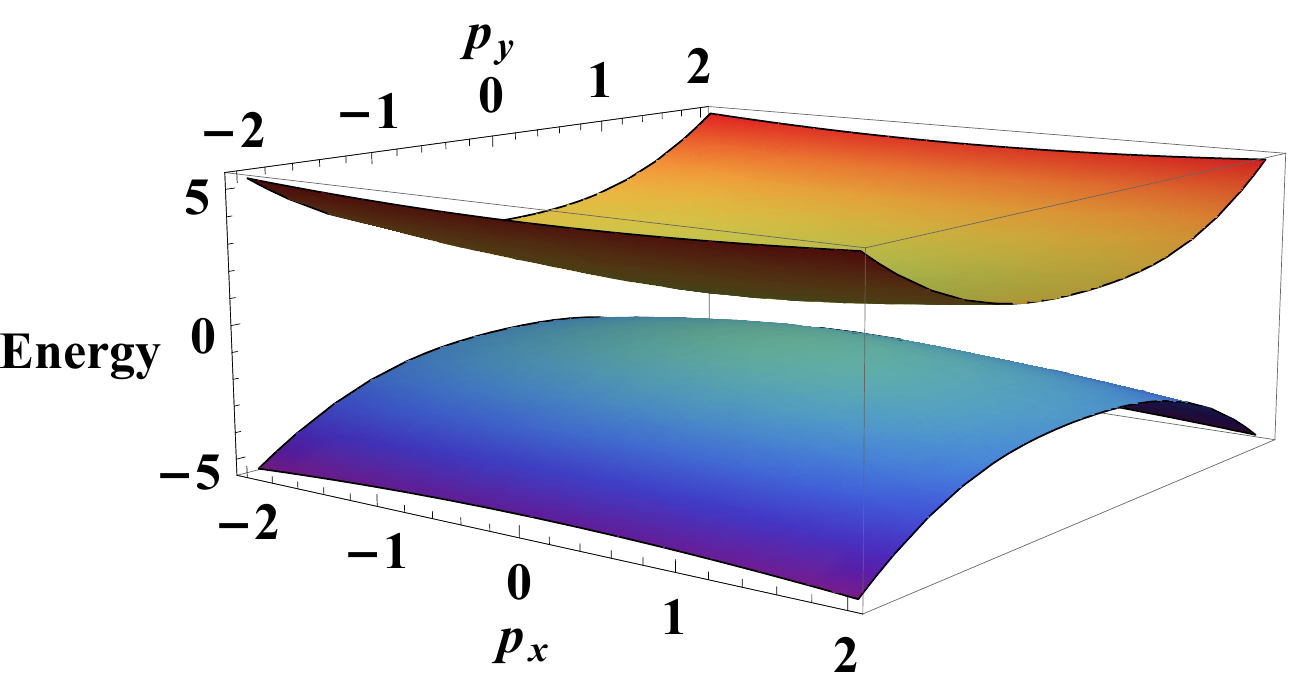}}
	\subfigure[]{\includegraphics[width=70mm,height=65mm]{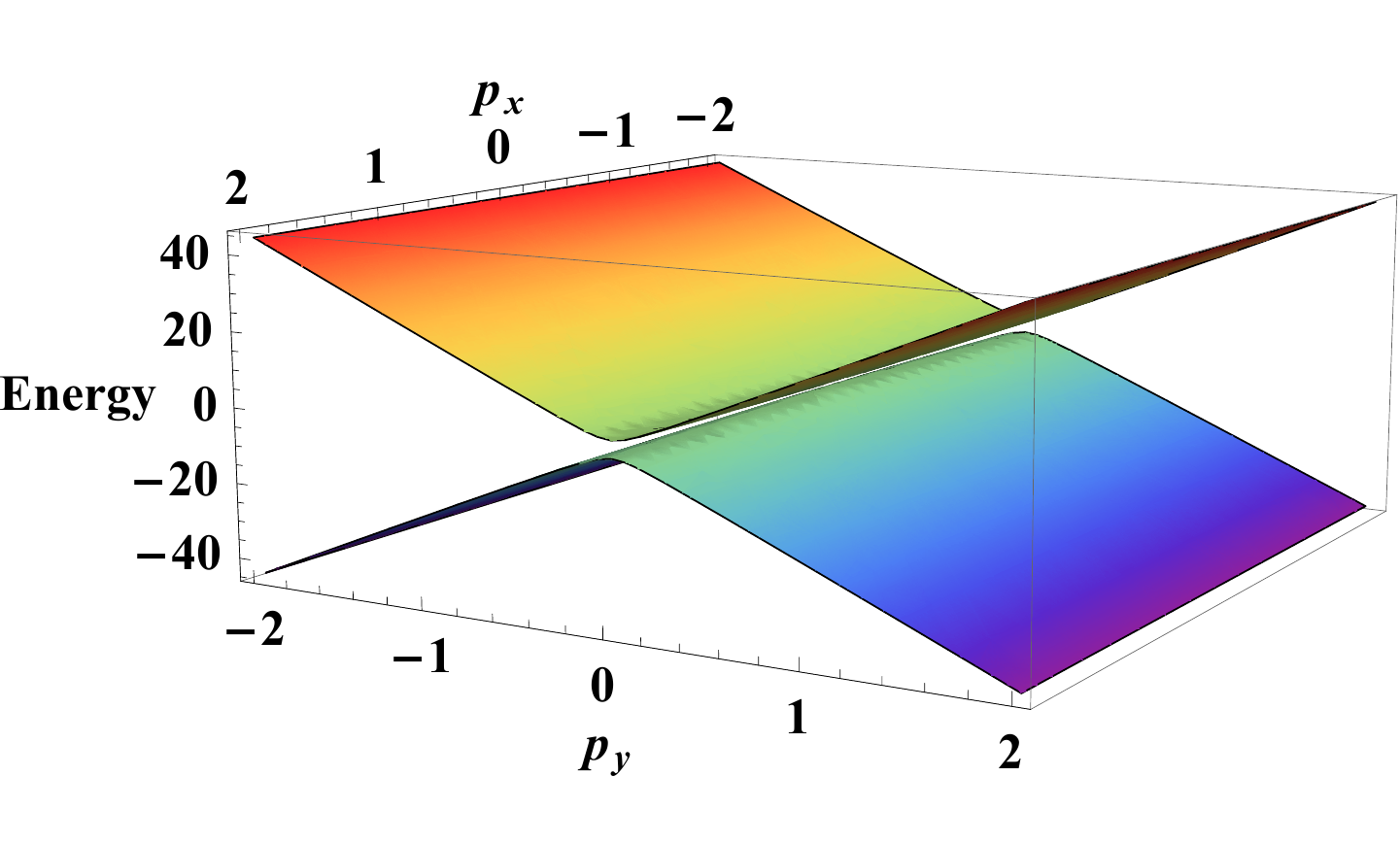}}\\
	\caption{\small (Color online) The 3D plot {\bf(a)} is showing energy vs momentum diagram without detuning i.e. in the absence of the external field (eigenvalues of $H_0$). The 3D plot {\bf(b)} is showing energy vs momentum, when detuning of the external field comes in the picture (eigenvalues of $H_{eff}$). For plotting, we have taken $\lambda=1$, $v_{F}=1$, $u=1$, $m=1$, $n=10$, and $\omega=10$  (parameters are plotted in unit of $\omega^{-1}$).}
	\label{tune}
	\label{3DPlot}
\end{figure}

Applying similar anology like earlier section (\ref{FloqP}),  the graphene Floquet equations, i.e.,
$i\frac{\partial}{\partial_{t}}|\phi(t)\rangle=H_{eff}|\phi(t)\rangle$ have form
\begin{subequations}
	\begin{align}i\hbar\frac{\partial}{\partial_t}\langle 0,1,n+1|\phi \rangle=\frac{(n+1) \lambda ^2}{\omega }\langle 0,1,n+1|\phi \rangle +v_F(p_{x}+i p_{y}) \langle 1,0,n+1|\phi \rangle
	\end{align}
	\begin{align}
	i\hbar\frac{\partial}{\partial_t}\langle 1,0,n+1|\phi \rangle=-\frac{(n+1) \lambda ^2}{\omega }\langle 1,0,n+1|\phi \rangle +v_F(p_{x}-i p_{y})\langle 0,1,n+1|\phi \rangle
	\end{align}
\end{subequations}
Choosing initial condition in quantized form like earlier section (\ref{RWA-appox}),
the population $P_{2G-Floq}(t)(=|\langle 1,0,n+1|\phi \rangle|^2)$ has polar form
\begin{align}
P_{2G-Floq}(t)=\sum_{n}\frac{\langle n \rangle^{n}}{n!}e^{-\langle n \rangle}\left[\frac{p^2 \omega ^2 v_F^2 \cos ^2\left[t \Omega_{\mathrm{Floq-G}}(n)\right]+\lambda ^4 (n+1)^2}{p^2 \omega ^2 v_F^2+\lambda ^4 (n+1)^2}\right].
\label{P2-FloqG}
\end{align}
The Floquet frequency $\Omega_{\mathrm{Floq-G}}(n)$ for graphene is defined as 
\begin{align}
\Omega_{\mathrm{Floq-G}}(n)=\sqrt{p^2  v_F^2+\frac{\lambda ^4 (n+1)^2}{\omega ^2}}
\label{ARWAG}
\end{align}
At Dirac point, the expression for $t_{\mathrm{colFloq}}$ and $t_{\mathrm{revFloq}}$ is
\begin{comment}
\[\Omega_{\mathrm{Floq-G}}(n)= \frac{\lambda ^2 (n+1)}{\omega }
\]
\[\Omega_{\mathrm{Floq-G}}(\bar{n})= \frac{\lambda ^2 \bar{n}}{\omega }
\]
\begin{align}
t_{\mathrm{col}}=\sqrt{\frac{2Log(10)}{\bar{n}}}\frac{\bar{n}}{\Omega_{\mathrm{Floq}}(\bar{n})},\quad t_{\mathrm{rev}}=\frac{2\pi\bar{n} }{\Omega_{\mathrm{Floq}}(\bar{n})}.
\end{align}
\begin{align}
t_{\mathrm{col}}=\sqrt{\frac{2Log(10)}{\bar{n}}}\frac{\bar{n} \omega }{\lambda ^2 \bar{n}},\quad t_{\mathrm{rev}}=\frac{2\pi\bar{n} \omega }{\lambda ^2 \bar{n}}.
\end{align}
\end{comment}
\begin{align}
t_{\mathrm{colFloq}}(\bar{n})=\sqrt{\frac{2Log(10)}{\bar{n}}}\frac{ \omega }{\lambda ^2 },\quad t_{\mathrm{revFloq}}(\bar{n})=\frac{2\pi \omega }{\lambda ^2 }.
\end{align}
The plot related with collapse-revival phenomenon of Floquet frequency of graphene is depicted in fig.\ref{collapse_revival_FloqG}{\bf(a)} and collapse-revival time in fig.\ref{collapse_revival_FloqG}{\bf(b)}. Here $t_{\mathrm{revFloq}}$ has constant nature but $t_{\mathrm{revFloq}}$ is varying with number of photon.
\newpage 
\noindent The plot showing anisotropic nature of Floquet frequency in phosphorene is depicted in fig.\ref{Rel}{\bf(a)} and compare between Floquet frequency of phosphorene [eq.(\ref{ARWAWP})] and graphene [eq.(\ref{ARWAG})] is depicted in fig.\ref{Rel}{\bf(b)}. The value of Floquet frequency of phosphorene is much larger than graphene at Dirac point (p=0), can be explicitly seen in fig.\ref{Rel}{\bf(b)}. Such threshold value of Floquet frequency comes due to the gap $m$ acting as the Dirac mass. 
\noindent The valence-conduction band tuning of phosphorene via Floquet frequency, can be seen in fig. (\ref{tune}). The band gap has been reduced, when external electromagnetic field has applied [fig.\ref{tune}{\bf(b)}]. Therefore, the Floquet frequency is a good tool for band tuning of 2D materials.     
\section{Quantum Bloch-Seigert shift}
\subsection{Phosphorene}
By using RWA, the expression for conventional Rabi frequency $\Omega_{\mathrm{RWA}}(n)$ [from section (\ref{RWAP})]  is 
\begin{align}
\Omega_{\mathrm{RWA-P}}(n)=\sqrt{\Delta_{p}^2+(n+1) \lambda^2\gamma}
\label{RWA}
\end{align}
Where $\Delta_{p}=(\omega-2\beta)$ is detuning parameter, $\beta$ and $\gamma$ by considering momentum vector in polar form i.e. $p_x=p \cos \theta $, $p_y=p \sin \theta$ and $\hbar=1$, is defined as 
\[\beta=\sqrt{\left(m+p^2 u \sin ^2\theta \right)^2+p^2 v_{F}^2 \cos ^2\theta }\]
\[\gamma =\frac{ \left[\left(m+p^2 u \sin ^2\theta\right)^2+p^4 u^2 \sin ^22 \theta\right]}{\left(m+p^2 u \sin ^2\theta\right)^2+p^2 v_{F}^2 \cos ^2\theta}
\]
The condition in RWA, the external frequency should equal to the frequency of two-level systems. Therefore, the counter-rotating terms are neglected \cite{haug2009quantum}. But RWA becomes invalid in the strong driving regimes, due to presence of counter-rotating terms. Therefore, a shift in conventional Rabi frequency resonance condition comes, known as Bloch-Siegert shift (BSS) \cite{bloch1940magnetic}. The expression of the conventional Rabi frequency with BSS   
\begin{align}
\Omega_{\mathrm{RWA-P-BSS}}(n)=\sqrt{\left(\Delta -\frac{\gamma  \lambda ^2 (n+1)}{8 \beta } \right)^2+ \lambda ^2 (n+1) \gamma-\left(\frac{\gamma  \lambda ^2 (n+1)}{8 \beta }\right)^2}
\label{RWABSS}
\end{align}
The term $\frac{\gamma  \lambda ^2 (n+1)}{8 \beta }$ is shift in conventional Rabi frequency, called as BSS . BSS is depending on the wave-vector angle $\theta$, so its nature becomes anisotropic in phosphorene.
\subsection{Graphene}
Similarly,  the expression for conventional Rabi frequency for graphene [from section (\ref{RWAG})] is 
\begin{align}
\Omega_{\mathrm{RWA-G}}(n)=\sqrt{\Delta_{G}^2+(n+1) \lambda^2}
\end{align}
Here $\Delta_{G}=\omega-2v_{F}\sqrt{p_{x}^2+p_{y}^2}$, which is detuning parameter in case of graphene. In presence of BSS, the conventional Rabi frequency have form 
\begin{align}
\Omega_{RWA-G-BSS} = \sqrt{ (n+1) \lambda^2 - \frac{ (n+1) \lambda^2 }{ 64 p^2  }
	+\left(\Delta - \frac{ (n+1) \lambda^2 }{ 8 p } \right)^2 }
\label{BSSG}
\end{align}
Therefore, isotropic nature of graphene conventional Rabi frequency BSS can be seen from above expression eq.(\ref{BSSG}). The term $\frac{ (n+1) \lambda^2 }{ 8 p }$  is shift in conventional Rabi frequency, called as BSS in graphene.
\begin{figure}[H]
	\centering 
	\subfigure[]{\includegraphics[width=70mm,height=65mm]{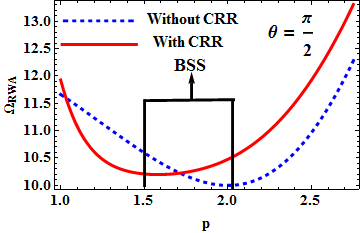}}
	\subfigure[]{\includegraphics[width=70mm,height=65mm]{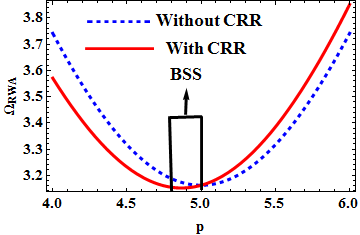}}\\
	\caption{ \small (Color online) Plot {\bf{(a)}} BSS in phosphorene corresponds to $\theta=\frac{\pi}{2}$. Plot {\bf{(b)}} BSS in graphene, independent on wave vector $\mathrm{\theta}$.  CRR is abbreviation of counter rotating term. For plotting, we considered, $\lambda=1$, $u=1$, $m=1$, $v_{F}=1$ $p=1$ $n=9$ and  $\omega=10$. Parameters are taken in unit of in the unit of $\lambda^{-1}$.}
	\label{BSS}
\end{figure}
\noindent In case of zero photon limit, i.e., $ n = 0 $, in eq.(\ref{RWA}) and eq.(\ref{RWABSS}), the Rabi oscillation is found. Such phenomenon is known as {\it {vacuum Rabi oscillation}}. The BSS in case of phosphorene is depicted in the  fig.\ref{BSS}{\bf({a})} and for graphene in the fig.\ref{BSS}{\bf({b})}. So BSS contains anisotropic nature in phosphorene.    

\section{Anisotropy of Quantum Floquet  Oscillations via Numerical Model in
	Phosphorene}
\begin{figure}[h]
	\centering
	\subfigure[]{\includegraphics[width=70mm,height=50mm]{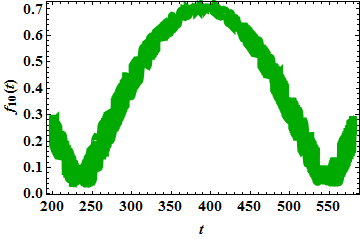}}
	\subfigure[]{\includegraphics[width=70mm,height=50mm]{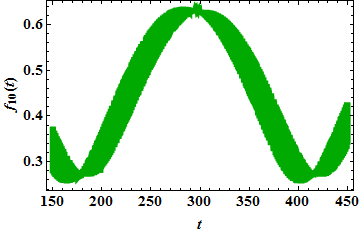}}\\
	\subfigure[]{\includegraphics[width=70mm,height=50mm]{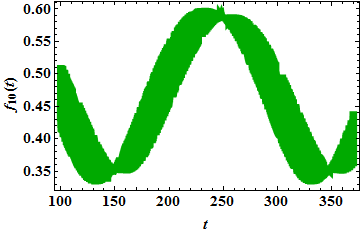}}
	\subfigure[]{\includegraphics[width=70mm,height=50mm]{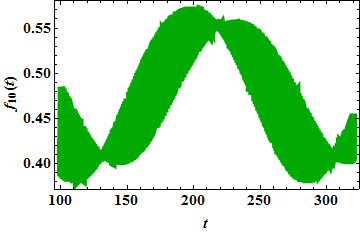}}\\
	\caption{\small (Color online) Time period (T) of oscillations in plot {\bf{(a)}} corresponds to Floquet frequency with a period of 312.6 {\bf{(b)}} corresponds to Floquet frequency with a period of 236.77 {\bf{(c)}} corresponds to Floquet frequency with a period of 198.25 {\bf{(d)}} corresponds to Floquet frequency with a period of 173.95. For plotting, we have taken $n=99$, $p=0.01$, $v_{ F}=1$, $u=1$, $m = .001$, $\lambda=1$ and $\omega=100$. Parameters are considered in the unit of $\lambda^{-1}$.}
	\label{numericalsolution}
\end{figure}

For explicitly justifying role of anisotropy in quantum Floquet oscillation of phosphorene,  the numerical solution of Floquet-Bloch equations have been described. From these equations, it can be seen clearly that wave-vector angle $\theta$ plays a major role in quantum Floquet oscillation of phosphorene. For phosphorene, the Floquet-Bloch equations are
\begin{subequations}
\begin{eqnarray}
i\hbar\frac{\partial}{\partial_t}\langle 0,1,n+1|\phi \rangle&=&
\left[\left\{ up^2_{y}+m\right\}+iv_{F} p_{x}\right] \mbox{  }\langle 1,0,n+1|\phi \rangle
+
e^{ -i \omega t}\left[ \left\{-i\frac{2up_{y}}{2v_{F}}\lambda  \mbox{   }\right\}-\frac{i}{2}\lambda \right]\;\sqrt{n+1} \mbox{  } \nonumber \\&&
\langle 1,0,n+1|\phi \rangle+e^{ i \omega t}
\left[\left\{i\frac{2up_{y}}{2v_{F}}\lambda \mbox{   }\right\}-\frac{i}{2}\lambda\right]\;\sqrt{n+1} \mbox{  }\langle 1,0,n+1|\phi \rangle
\label{Floq-Bloch1}
\end{eqnarray}
\begin{eqnarray}
i\hbar\frac{\partial}{\partial_t}\langle 1,0,n+1|\phi \rangle &=&
\left[ \left\{ \left(up^2_{y}\right)+m\right\}-iv_{F} p_{x}\right] \mbox{  }\mbox{  } \langle 0,1,n+1|\phi \rangle
+e^{ -i \omega t} \left[ \left\{-i\frac{2up_{y}}{2v_{F}}\lambda \right\}+\frac{i}{2}\lambda  \right]\;\sqrt{n+1} \mbox{  } \nonumber \\&&
\langle 0,1,n+1|\phi \rangle+e^{ i \omega t} \left[i\left\{\frac{2up_{y}}{2v_{F}}\lambda \right\}+\frac{i}{2}\lambda\right]\mbox{  }\sqrt{n+1} \mbox{  }\langle 0,1,n+1|\phi \rangle
\label{Floq-Bloch2}
\end{eqnarray}
\end{subequations}
\noindent Where we have taken $n\approx n+1$ and $\lambda=1$. By using NDSolve routine in Mathematica software \cite{Mathematica}, the eq.(\ref{Floq-Bloch1}) and eq.(\ref{Floq-Bloch2}) solved numerically and shown in fig.(\ref{numericalsolution}). The time period belongs to Floquet frequencies is described in table (\ref{numericvalue}), which is showing anisotropic nature and verifying the time period of analytical Floquet frequency (eq.(\ref{ARWAWP})). As we increase value of $p$, the time period of Floquet oscillations decreasing continuously irrespective of the value of $\theta$ (table (\ref{numericvalue})). Therefore, it can be said that, anisotropy is a crucial and significant parameter for Floquet oscillations in phosphorene. 
\FloatBarrier

\begin{table}[H]
	\begin{center}
		\begin{tabular}{|c|c|c|c|c|c|}
			\hline
			&&&&\\
			$\huge p$ &  T$|_{\theta =0 }$   &  T$|_{\theta = \pi/6 }$ &  T$|_{\theta =\pi/4 }$ &  T$|_{\theta = \pi/3}$   \\
			%  & $\theta=0$  &  $\theta=\frac{\pi}{4}$& $\theta=\frac{\pi}{3}$ &$\theta=\frac{\pi}{2}$ \\
			%  &&&&\\
			\hline
			$ 0.01$  &312.6 &236.77& 198.25&173.95\\
			\hline
			$ 0.02$ &156.88 &118.64& 99.27&87.07\\
			\hline
			$ 0.03  $  &104.66 &79.12& 66.19&58.06\\
			\hline
			$ 0.04$ &78.51 &59.35& 49.65&43.54\\
			\hline
			0.05 &62.82 &47.48& 39.72&34.83\\
			\hline
		\end{tabular}
		\caption{\small Time period (T) of oscillations for various values of $\mathrm{\theta}$, confirming the anisotropic behaviour of Floquet oscillations in phashphorene. T is in the  units of $\lambda^{-1}$.}
		\label{numericvalue}
	\end{center}
\end{table}
\section{Conclusions}
We have described, how {\it anisotropy} is playing major role in Floquet theory and rotating wave approximation in case of phosphorene. Floquet theory becomes important in case of low energy physics, so in the low energy region, the band tuning of phosphorene can be done via Floquet frequency. There is no role of anisotropy in case graphene. The results of Floquet frequency and Rabi frequency are compared between graphene and phosphorene, wherever required. But, Floquet theory becomes more important in case of low energy physics. Therefore, Floquet theory is an important tool for the study of the 2D materials.        
\bibliographystyle{spphys}
\bibliography{Phosphorene}
\end{document}